\documentstyle[11pt]{article} 
\input{pictex.sty}
\newcommand{\g}{\gamma}
\newcommand{\prop}{\Delta}

\newcommand{\dsl}{\not \! \partial}
\renewcommand{\d}{\partial}
\newcommand{\T}{{\mathrm T}}
\newcommand{\B}{{\mathrm B}}

\renewcommand{\O}{{\mathcal O}}
\begin{document}
{\hfill \parbox{6cm}{\begin{center} 
				    UG-FT-81/97
		      \end{center}}} 
	      
\vspace*{1cm}                               
\begin{center}
\large{\bf Constrained differential 
renormalization\footnote{Presented at the XXI International School of
Theoretical Physics ``Recent progress in theory and phenomenology
of fundamental interactions'', Ustro\'n, Poland, September 19-24,
1997.}} 
\vskip .3truein
\centerline {\sc F. del \'Aguila and M. P\'erez-Victoria}
\vspace{0.5cm}
\centerline {\it Dpto. de F\'{\i}sica Te\'orica y del Cosmos,} 
\centerline {\it Universidad de Granada, 18071 Granada, Spain}
\end{center}
\vspace{0.5cm}
\noindent We review the method of differential renormalization, 
paying special attention to a
new constrained version for symmetric theories.
  
\section{Introduction}
Popular regularization and renormalization methods work in momentum 
space. Typically, the divergences which appear in loop integrals at
large internal momenta (ultraviolet divergences) are first 
regulated---i.e., the integrals are modified so that they are finite,
but diverge in the limit of no regulator---and then substracted by
adding the necessary counterterms. For example, in dimensional
regularization~\cite{dimreg} the regulated integrals are defined 
in $n$ dimensions by analytical continuation, 
with $n$ an arbitrary complex number. 
Eventually, the poles appearing at $n=4$ 
are cancelled by appropriate counterterms and a finite (renormalized) 
result is obtained for $n \rightarrow 4$. 
Renormalization without intermediate
regularization is also possible, as in the BPHZ 
method~\cite{BPHZ}, where the
first terms of the Taylor expansion in external momenta of
the integrand are
substracted off before integrating. Although momentum space is more
natural for calculations of scattering amplitudes with fixed external
momenta, nothing prevents us from working in coordinate space and, if
required, perform a Fourier transform at the end. In coordinate space,
ultraviolet divergences correspond to a singular behaviour at short
distances.

In Ref.~\cite{FJL} a method of renormalization in coordinate space 
was proposed: differential renormalization (DR). 
It is based on the observation that primitively divergent
Feynman graphs are well defined in coordinate space for non-coincident
points, but too singular at coincident points to allow for Fourier
transform. In other words, the corresponding expressions are not
well-behaved distributions. The idea of DR is to substitute the
singular expressions by derivatives of well-behaved distributions, in
such a way that the former (`bare') and latter 
(`renormalized') expressions are equal almost
everywhere. These derivatives are prescribed to act formally by
parts in integrals such as Fourier transforms. In this
way, finite Green functions are obtained, without the
need of intermediate regularization. DR acts directly on
bare Feynman graphs and does not introduce explicit counterterms.
\begin{figure}
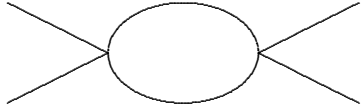

\font\thinlinefont=cmr5
\begingroup\makeatletter\ifx\SetFigFont\undefined%
\gdef\SetFigFont#1#2#3#4#5{%
  \reset@font\fontsize{#1}{#2pt}%
  \fontfamily{#3}\fontseries{#4}\fontshape{#5}%
  \selectfont}%
\fi\endgroup%
\mbox{\beginpicture
\setcoordinatesystem units <0.7cm,0.7cm>
\unitlength=1.04987cm
\linethickness=1pt
\setplotsymbol ({\makebox(0,0)[l]{\tencirc\symbol{'160}}})
\setshadesymbol ({\thinlinefont .})
\setlinear
%
%
\linethickness= 0.500pt
\setplotsymbol ({\thinlinefont .})
\ellipticalarc axes ratio  1.429:0.953  360 degrees 
	from 14.287 17.145 center at 12.859 17.145
%
%
\linethickness= 0.500pt
\setplotsymbol ({\thinlinefont .})
\plot  9.525 18.098 11.430 17.145 /
\plot 11.430 17.145  9.525 16.192 /
%
%
\linethickness= 0.500pt
\setplotsymbol ({\thinlinefont .})
\plot 16.192 18.098 14.287 17.145 /
\plot 14.287 17.145 16.192 16.192 /
\linethickness=0pt
\putrectangle corners at 5.500 15.123 and 16.218 18.167
\endpicture}
\caption{One-loop diagram contributing to the 
four-point vertex in $\Phi^4$. \label{figphi}}
\end{figure}
The procedure is
best illustrated in terms of one example: the one-loop four-point bubble
graph of massless $\Phi^4$ (Fig.~\ref{figphi}). 
We work in euclidean
space, which leads to simpler functions.
The massless
propagator in position space is $\prop(x-y)=\frac{1}{4\pi^2} 
\frac{1}{(x-y)^2}$
and the vertex, $-\lambda \delta(x_1-x_4) \delta(x_2-x_4) \delta(x_3-x_4)$.
The bare expression for the amputated graph is
\begin{equation}
  \Gamma(x_1,x_2,x_3,x_4)  =  \frac{\lambda^2}{2} \frac{1}{16\pi^4}
  \delta(x_1-x_2)\delta(x_3-x_4) \frac{1}{(x_1-x_3)^4 } + {\mathrm 
  2~perms.}
\label{barephi}
\end{equation}
This involves
the singular function $\frac{1}{x^4}$, which has a logarithmically 
divergent Fourier transform. 
To renormalize it with DR, one must solve a differential 
equation and find $f(x^2)$ such that 
\begin{equation}
  \frac{1}{x^4} = \Box f(x^2)
\label{diffeq} 
\end{equation}
for $x \not = 0$. Actually, one derivative would be enough in
this case,
but one uses the D'Alambertian $\Box=\d_\mu \d_\mu$ to 
preserve manifest euclidean invariance. 
The solution of Eq.~(\ref{diffeq}) is 
\begin{equation}
  f(x^2)= - \frac{1}{4} \frac{\log x^2 M^2}{x^2} ~,
\end{equation}
where $M$ is an arbitrary constant 
with dimensions of mass,  
required for dimensional reasons, and we have omitted a possible
but irrelevant additive constant. Although we shall not
discuss it here, it is worth mentioning that the constant
$M$ plays a central role in DR: the renormalized amplitudes satisfy
renormalization group equations, with $M$ the 
renormalization scale.
The renormalized expression of the singular function reads
\begin{equation}
  \left[ \frac{1}{x^4} \right]^R = - \frac{1}{4} \Box
  \frac{\log x^2 M^2}{x^2} ~.
\label{DRidentity} 
\end{equation}
With the formal integration by parts rule, this is a tempered
distribution which admits a finite Fourier transform:
\begin{eqnarray}
  \int {\mathrm d}^4 \!x \, e^{i p \cdot x} 
  \left[ \frac{1}{x^4} \right]^R & = &
  - p^2  \int {\mathrm d}^4 \!x \, e^{i p \cdot x} 
  (-\frac{1}{4}) \frac{\log x^2 M^2}{x^2}  \nonumber \\
  & = &- \pi^2 \log (\frac{p^2}{\bar{M}^2}) ~,
\label{DRFT}
\end{eqnarray}
where $\bar{M} = 2M/\gamma_E$,  and $\gamma_E=1.781...$ is  Euler's
constant.
Substituting Eq.~(\ref{DRidentity}) in Eq.~(\ref{barephi}), the 
renormalized vertex
graph is obtained:
\begin{eqnarray}
  \Gamma^R(x_1,x_2,x_3,x_4) & = & - \frac{\lambda^2}{128\pi^4} 
  \delta(x_1-x_2)\delta(x_3-x_4) \Box 
  \frac{\log (x_1-x_3)^2 M^2}{(x_1-x_3)^2} \nonumber \\
  && \mbox{} + {\mathrm 2~perms.}
\end{eqnarray}
The renormalized expression in momentum space follows  
from Eq.~(\ref{DRFT}):
\begin{eqnarray}
  \hat{\Gamma}^R(p_1,p_2,p_3,p_4) & = &
  - \frac{\lambda^2}{32\pi^2} \log \left[\frac{(q_1+q_2)^2
  (q_2+q_3)^2 (q_1+q_3)^2}{\bar{M}^6} \right] \nonumber \\
  && \mbox{} \times (2\pi)^4 \delta(\sum_i q_i).
\end{eqnarray}

DR has been successfully applied in different contexts:
the Wess-Zumino model~\cite{WZ}, 
lower-dimensional~\cite{Ramon} and non-abelian
gauge theories~\cite{nonabelian}, two-loop 
QED~\cite{QED}, a chiral model~\cite{chiral}, a 
non-relativistic anyon model~\cite{anyon}, 
curved space-time and finite temperature~\cite{curved}, 
the calculation of $(g-2)_l$ in supergravity~\cite{g2}, 
Chern-Simons theories~\cite{coreanos} and non-perturbative
calculations in supersymmetric gauge 
theories~\cite{nonperturbative}. Other formal aspects of the
method have been developed in 
Refs. \cite{counterterm,systematic,Nuria,massiveDR} and 
different versions of DR can be found in~\cite{Smirnov,Schnetz}.
 
When symmetries are an issue, it is important that
the renormalization program preserves the
corresponding Ward identities. In general, even when
the regularization procedure breaks some relevant symmetry,
one can still recover it by the addition of finite
counterterms\footnote{The
exception is called an anomaly: the
quantum renormalized theory does not have a symmetry
of the classical theory.}. In practical calculations and formal
proofs to all orders, it is nevertheless more convenient
to have a method that directly preserves the
Ward identities. The great success of dimensional
regularization is mainly due to the fact that it automatically
respects gauge invariance. It is known, however, that it has
problems in dimension-dependent theories like chiral
and supersymmetric theories. Its variant
dimensional reduction~\cite{dimred} is usually employed in
these cases, although inconsistencies may arise at high 
orders~\cite{dimredprob}.
DR does not change the space-time
dimension and it was expected to become a renormalization
procedure respecting gauge and chiral
symmetry. In its original version, however, this
is not automatic. The ambiguities inherent
to the manipulation of
singular functions are taken care of by introducing
arbitrary renormalization scales for different diagrams. 
Different choices of the renormalization scales
give rise to different renormalization schemes and 
only a subset of these schemes corresponds to a symmetric
renormalized theory. Hence,
the scales must be fixed to enforce the relevant
Ward identities. A change of renormalization scales is equivalent
to the addition of finite counterterms, so the situation
does not differ much from the one with symmetry-breaking
regulators.

In Refs.~\cite{g2,CDR} a procedure of DR was proposed which
fixes all the manipulations and only introduces the
necessary renormalization group scale. This {\em constrained}
DR has been explicitly shown
to respect the one-loop Ward identities of abelian
gauge symmetry~\cite{CDR}\footnote{In Ref.~\cite{gaugeSmirnov}
Smirnov presented an abelian gauge invariant method within
his version of DR.} and to preserve supersymmetry
in a supergravity calculation~\cite{g2}.

In the following, we first describe the method of 
constrained DR and then use it to
calculate in detail the electron 
self-energy  and the vertex correction in QED, as an illustration. 
We also derive the corresponding momentum space
expressions and
check that the corresponding  Ward identity is 
automatically fulfilled. 
 

\section{Constrained differential renormalization}
In this section we briefly describe the  constrained procedure of 
DR at one loop introduced in Ref.~\cite{CDR}. The idea is to find a
consistent way of performing the manipulations of singular
expressions carried out in the process of renormalization.
It turns out that a small set of formal rules is sufficient to 
completely fix  the renormalization
scales (except one, associated with the renormalization group
invariance). Furthermore, the resulting renormalized
amplitudes were explicitly shown in Refs.~\cite{CDR,g2}
to satisfy the one-loop Ward identities of abelian gauge symmetry
and to render a vanishing value for the magnetic moment of a 
charged lepton
in supergravity (which is the required value if supersymmetry
is respected~\cite{Ferrara}). 

The set of rules contains the two basic DR rules: the use of DR 
identities like Eq.~(\ref{DRidentity}) (always with the same 
renormalization scale!) ({\em rule 1}) and the formal integration by
parts prescription ({\em rule 2}). In addition, we need another two
rules.
One is ({\em rule 3}): 
\begin{equation}
  [F(x,x_1,...,x_n) \delta(x-y)]^R =  [F(x,x_1,...,x_n)]^R
  \delta(x-y)~,
\end{equation}
where $F$ is an arbitrary function. The other one 
requires
the general validity of the propagator equation
({\em rule 4}):
\begin{equation} 
  F(x,x_1,...,x_n) \Box \prop(x) = 
  F(x,x_1,...,x_n) (- \delta(x)) ~,
  \label{propeq}
\end{equation}
where $\prop(x) = \frac{1}{4\pi^2} \frac{1}{x^2}$ is
the massless propagator (the massive case is
analogous).
This is a valid mathematical identity between tempered
distributions if F is well-behaved enough. 
This rule formally extends its range of applicability 
to an arbitrary function.
The main point of the constrained method is to 
require consistency
of renormalization with these rules. Such requirement 
fixes all the ambiguities in DR at least to one loop.
Let us explain how the rules are used in practice 
with some simple examples. 
The results will be used in the next section. 
We first introduce some convenient notation: we
define the {\em bubble} and {\em triangular} basic functions as
\begin{eqnarray}
  \B[\O] & \equiv & \prop(x) \O \prop(x) ~, \label{B}\\
  \T[\O] & \equiv & \prop(x) \prop(y) \O^x \prop(x-y)~, \label{T} 
\end{eqnarray}
where $\O$ is a differential operator. The significance of
this kind of functions is that any one-loop bubble or triangular
Feynman diagram 
can be expressed in terms of them (and their
derivatives) using only algebraic
manipulations and the Leibnitz rule for derivatives.
Hence, the problem of renormalization reduces (at this
order) to finding the renormalized expressions of
the singular basic functions. 
Note that $\B[\O]$ is singular\footnote{In this paper the term 
`singular' should always be undertood as `too singular to allow for
a Fourier transform'.}
at $x=0$ for any $\O$,
whereas $\T[\O]$ is only singular (at $x=y=0$) when
$\O$ contains two or more derivatives.
These basic functions are easily renormalized using
the set of rules described above. For example,
\begin{eqnarray}
  \B^R[1] & = & \left[ \prop(x) \prop(x) \right]^R \nonumber \\
     & = & \frac{1}{(4\pi^2)^2} 
           \left[ \frac{1}{x^4} \right]^R \nonumber \\
     & \stackrel{\mathrm Rules~1,2}{=} & 
           -\frac{1}{4} \frac{1}{(4\pi^2)^2} 
           \Box \frac{\log x^2 M^2}{x^2}
\label{renB}
\end{eqnarray}
and
\begin{eqnarray}
\T^R[\Box]  & = & \left[ \prop(x) \prop(y) \Box^x \prop(x-y) 
                  \right]^R \nonumber \\
   & \stackrel{\mathrm Rule~4}{=} & 
         - \left[ (\prop(x))^2 \delta(x-y) \right]^R \nonumber \\
   & = & - \left[ \B[1](x) \delta(x-y) \right]^R \nonumber \\
   & \stackrel{\mathrm Rule~3}{=} & 
         - \B^R[1](x) \delta(x-y) \nonumber \\
   & \stackrel{\mathrm Eq.~(\ref{renB})}{=} &  
         \frac{1}{4} \frac{1}{(4\pi^2)^2} \Box
         \frac{\log x^2 M^2}{x^2} \delta(x-y) ~. 
\label{renTbox}
\end{eqnarray}
For basic functions with non-trivial tensor structure
the procedure is more involved and can be found
in Ref.~\cite{CDR}. 
In particular,  we shall need in the following the identity:
\begin{eqnarray}
\T^R[\d_\mu\d_\nu] & = & 
    \T[\d_\mu\d_\nu - \frac{1}{4} 
    \delta_{\mu\nu}\Box] \nonumber \\
    && \mbox{} + (\frac{1}{16} \frac{1}{(4\pi^2)^2} 
    \Box \frac{\log x^2 M^2}{x^2} \delta(x-y) 
    -  \frac{1}{32} \frac{1}{4\pi^2} 
    \delta(x) \delta(y) ) \delta_{\mu\nu}
\label{renTdd}
\end{eqnarray}
where the first term is finite thanks to its tracelessness.
The local term proportional to $\delta(x) \delta(y)$
was not considered in the earlier literature (before Ref.~\cite{CDR},
where Eq.~(\ref{renTdd}) is worked out in detail) 
and comes from imposing consistency with the propagator equation
(rule~4).
Notice that
\begin{equation}
   \delta_{\mu\nu} \T^R[\d_\mu\d_\nu] \not =
   \left[ \delta_{\mu\nu} \T[\d_\mu\d_\nu] \right]^R ~.
\end{equation}
This might seem strange, but in fact also occurs in
other schemes like dimensional regularization
or Pauli-Villars. Differentially renormalized expressions
of basic functions appearing in one-, two- and three-point
one-loop Green functions can be found in the
tables of Ref.~\cite{CDR}.
The treatment of four-point one-loop Green functions
will be presented in Ref.~\cite{SQED}. 


\section{Simple applications}
\begin{figure}
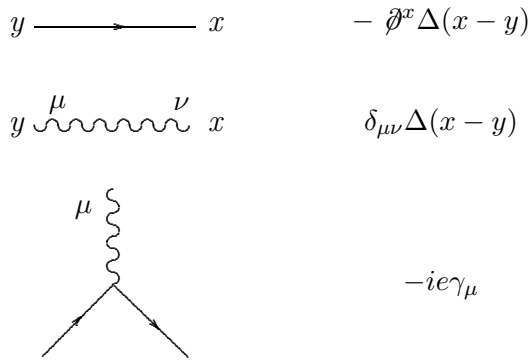

\font\thinlinefont=cmr5
\begingroup\makeatletter\ifx\SetFigFont\undefined%
\gdef\SetFigFont#1#2#3#4#5{%
  \reset@font\fontsize{#1}{#2pt}%
  \fontfamily{#3}\fontseries{#4}\fontshape{#5}%
  \selectfont}%
\fi\endgroup%
\mbox{\beginpicture
\setcoordinatesystem units <0.5cm,0.5cm>
\unitlength=1.04987cm
\linethickness=1pt
\setplotsymbol ({\makebox(0,0)[l]{\tencirc\symbol{'160}}})
\setshadesymbol ({\thinlinefont .})
\setlinear
%
%
\linethickness= 0.500pt
\setplotsymbol ({\thinlinefont .})
\circulararc 180.769 degrees from  6.824 20.240 center at  6.982 20.239
%
%
\linethickness= 0.500pt
\setplotsymbol ({\thinlinefont .})
\circulararc 180.000 degrees from  7.457 20.240 center at  7.298 20.240
%
%
\linethickness= 0.500pt
\setplotsymbol ({\thinlinefont .})
\circulararc 180.769 degrees from  7.457 20.240 center at  7.615 20.239
%
%
\linethickness= 0.500pt
\setplotsymbol ({\thinlinefont .})
\circulararc 180.000 degrees from  8.090 20.240 center at  7.931 20.240
%
%
\linethickness= 0.500pt
\setplotsymbol ({\thinlinefont .})
\circulararc 180.769 degrees from  8.090 20.240 center at  8.248 20.239
%
%
\linethickness= 0.500pt
\setplotsymbol ({\thinlinefont .})
\circulararc 180.769 degrees from  8.721 20.240 center at  8.563 20.241
%
%
\linethickness= 0.500pt
\setplotsymbol ({\thinlinefont .})
\circulararc 180.000 degrees from  8.721 20.240 center at  8.879 20.240
%
%
\linethickness= 0.500pt
\setplotsymbol ({\thinlinefont .})
\circulararc 180.769 degrees from  9.354 20.240 center at  9.196 20.241
%
%
\linethickness= 0.500pt
\setplotsymbol ({\thinlinefont .})
\circulararc 180.000 degrees from  9.354 20.240 center at  9.512 20.240
%
%
\linethickness= 0.500pt
\setplotsymbol ({\thinlinefont .})
\circulararc 180.769 degrees from  9.986 20.240 center at  9.829 20.241
%
%
\linethickness= 0.500pt
\setplotsymbol ({\thinlinefont .})
\circulararc 180.000 degrees from  6.824 20.240 center at  6.665 20.240
%
%
\linethickness= 0.500pt
\setplotsymbol ({\thinlinefont .})
\circulararc 180.769 degrees from  6.191 20.240 center at  6.349 20.239
%
%
\linethickness= 0.500pt
\setplotsymbol ({\thinlinefont .})
\circulararc 180.000 degrees from  9.986 20.240 center at 10.145 20.240
%
%
\linethickness= 0.500pt
\setplotsymbol ({\thinlinefont .})
\circulararc 181.538 degrees from  8.278 17.913 center at  8.276 17.757
%
%
\linethickness= 0.500pt
\setplotsymbol ({\thinlinefont .})
\circulararc 180.000 degrees from  8.278 17.280 center at  8.278 17.439
%
%
\linethickness= 0.500pt
\setplotsymbol ({\thinlinefont .})
\circulararc 181.538 degrees from  8.278 17.280 center at  8.276 17.124
%
%
\linethickness= 0.500pt
\setplotsymbol ({\thinlinefont .})
\circulararc 180.000 degrees from  8.278 16.648 center at  8.278 16.806
%
%
\linethickness= 0.500pt
\setplotsymbol ({\thinlinefont .})
\circulararc 181.538 degrees from  8.278 16.648 center at  8.276 16.491
%
%
\linethickness= 0.500pt
\setplotsymbol ({\thinlinefont .})
\circulararc 180.000 degrees from  8.278 16.017 center at  8.278 16.176
%
%
\linethickness= 0.500pt
\setplotsymbol ({\thinlinefont .})
\circulararc 180.000 degrees from  8.278 17.913 center at  8.278 18.072
%
%
\linethickness= 0.500pt
\setplotsymbol ({\thinlinefont .})
\circulararc 181.538 degrees from  8.278 18.546 center at  8.276 18.390
%
%
\linethickness= 0.500pt
\setplotsymbol ({\thinlinefont .})
\plot  8.272 15.987  6.413 14.097 /
%
%
\linethickness= 0.500pt
\setplotsymbol ({\thinlinefont .})
\plot  8.319 15.970 10.270 14.082 /
%
%
\linethickness= 0.500pt
\setplotsymbol ({\thinlinefont .})
\plot  9.377 14.944  9.493 14.827 /
%
%
\plot  9.269 14.962  9.493 14.827  9.359 15.052 /
%
%
%
\linethickness= 0.500pt
\setplotsymbol ({\thinlinefont .})
\plot  7.303 15.007  7.429 15.145 /
%
%
\plot  7.304 14.915  7.429 15.145  7.211 15.001 /
%
%
%
\linethickness= 0.500pt
\setplotsymbol ({\thinlinefont .})
\putrule from  6.191 22.860 to  8.572 22.860
%
%
\plot  8.319 22.796  8.572 22.860  8.319 22.924 /
%
%
%
\linethickness= 0.500pt
\setplotsymbol ({\thinlinefont .})
\putrule from  8.572 22.860 to 10.478 22.860
%
%
\put{\SetFigFont{12}{14.4}{\rmdefault}{\mddefault}{\itdefault}y} [lB] 
at  5.5 20.098
%
%
%
\put{\SetFigFont{12}{14.4}{\rmdefault}{\mddefault}{\itdefault}x} [lB] 
at 10.8 20.098
%
%
%
\put{$\mu$} at 6.8 20.8
\put{$\nu$} at 10.1 20.8
\put{\SetFigFont{12}{14.4}{\rmdefault}{\mddefault}{\itdefault}y} [lB] 
at  5.5 22.765
%
%
\put{\SetFigFont{12}{14.4}{\rmdefault}{\mddefault}{\itdefault}x} [lB] 
at 10.8 22.765
%
%
\put{$\mu$} at 7.5 18
%
%
\put{$-\dsl^x \prop(x-y)$} at 17 23.0
\put{$\delta_{\mu\nu} \prop(x-y)$} at 17 20.3
\put{$-i e \g_\mu$} at 17 16.0
\linethickness=0pt
\putrectangle corners at  2 22.993 and 6.954 13.0
\endpicture}
\caption{Feynman rules of massless QED
in euclidean coordinate space. \label{Feynrules}}
\end{figure}

We now show how the method works in practice with
two detailed examples:
the renormalization of the one-loop 
1PI electron self-energy and electron-electron-photon vertex
in massless QED.
The Feynman rules of massless QED in euclidean coordinate
space are gathered in Fig.~\ref{Feynrules}, with
$\{ \g_\mu , \g_\nu \} = 2 \delta_{\mu\nu}$.
We work in the Feynman gauge. The case of an arbitrary Lorentz
gauge was discussed in Ref.~\cite{CDR}.
\begin{figure}
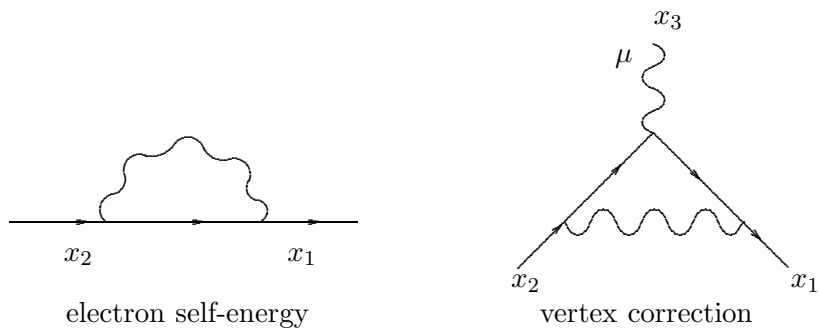

\begin{center}
\font\thinlinefont=cmr5
\begingroup\makeatletter\ifx\SetFigFont\undefined%
\gdef\SetFigFont#1#2#3#4#5{%
  \reset@font\fontsize{#1}{#2pt}%
  \fontfamily{#3}\fontseries{#4}\fontshape{#5}%
  \selectfont}%
\fi\endgroup%
\mbox{\beginpicture
\setcoordinatesystem units <0.5cm,0.5cm>
\unitlength=1.04987cm
\linethickness=1pt
\setplotsymbol ({\makebox(0,0)[l]{\tencirc\symbol{'160}}})
\setshadesymbol ({\thinlinefont .})
\setlinear
%
%
\linethickness= 0.500pt
\setplotsymbol ({\thinlinefont .})
\putrule from  1.901 19.329 to  3.954 19.329
%
%
\plot  3.700 19.266  3.954 19.329  3.700 19.393 /
%
%
%
\linethickness= 0.500pt
\setplotsymbol ({\thinlinefont .})
\putrule from  3.954 19.329 to  7.036 19.329
%
%
\plot  6.782 19.266  7.036 19.329  6.782 19.393 /
%
%
%
\linethickness= 0.500pt
\setplotsymbol ({\thinlinefont .})
\putrule from 10.118 19.329 to 11.144 19.329
%
%
\linethickness= 0.500pt
\setplotsymbol ({\thinlinefont .})
\putrule from  7.036 19.329 to 10.118 19.329
%
%
\plot  9.864 19.266 10.118 19.329  9.864 19.393 /
%
%
%
\linethickness= 0.500pt
\setplotsymbol ({\thinlinefont .})
\circulararc 108.924 degrees from  4.600 20.055 center at  4.561 20.469
%
%
\linethickness= 0.500pt
\setplotsymbol ({\thinlinefont .})
\circulararc 117.940 degrees from  5.508 21.101 center at  5.358 20.707
%
%
\linethickness= 0.500pt
\setplotsymbol ({\thinlinefont .})
\circulararc 62.820 degrees from  5.489 21.101 center at  5.671 21.837
%
%
\linethickness= 0.500pt
\setplotsymbol ({\thinlinefont .})
\circulararc 129.775 degrees from  7.063 21.306 center at  6.645 21.130
%
%
\linethickness= 0.500pt
\setplotsymbol ({\thinlinefont .})
\circulararc 81.207 degrees from  7.063 21.306 center at  7.586 21.552
%
%
\linethickness= 0.500pt
\setplotsymbol ({\thinlinefont .})
\circulararc 127.342 degrees from  8.278 20.449 center at  7.878 20.592
%
%
\linethickness= 0.500pt
\setplotsymbol ({\thinlinefont .})
\circulararc 130.713 degrees from  8.278 20.449 center at  8.541 20.238
%
%
\linethickness= 0.500pt
\setplotsymbol ({\thinlinefont .})
\circulararc 130.371 degrees from  8.619 19.372 center at  8.463 19.621
%
%
\linethickness= 0.500pt
\setplotsymbol ({\thinlinefont .})
\circulararc 121.495 degrees from  4.600 20.055 center at  4.718 19.660
%
%
\put{$x_1$} [lB] at  9.296 18.303
%
%
\put{$x_2$} [lB] at  3.338 18.303
\linethickness= 0.500pt
\setplotsymbol ({\thinlinefont .})
%
%
\plot 19.020 24.066 	19.098 24.030
	19.162 23.995
	19.254 23.926
	19.304 23.853
	19.319 23.768
	19.293 23.662
	19.226 23.583
	19.131 23.523
	19.021 23.471
	18.912 23.419
	18.817 23.359
	18.750 23.280
	18.724 23.173
	18.750 23.067
	18.817 22.988
	18.912 22.928
	19.021 22.876
	19.131 22.824
	19.226 22.763
	19.293 22.685
	19.319 22.578
	19.293 22.472
	19.226 22.393
	19.131 22.332
	19.021 22.281
	18.912 22.229
	18.817 22.168
	18.750 22.090
	18.724 21.984
	18.739 21.899
	18.789 21.826
	18.880 21.757
	18.944 21.722
	19.020 21.685
	/
%
%
\linethickness= 0.500pt
\setplotsymbol ({\thinlinefont .})
\plot 15.416 18.095 16.565 19.249 /
%
%
\plot 16.431 19.024 16.565 19.249 16.341 19.114 /
%
%
%
\linethickness= 0.500pt
\setplotsymbol ({\thinlinefont .})
\plot 16.565 19.249 18.144 20.828 /
%
%
\plot 18.009 20.603 18.144 20.828 17.920 20.693 /
%
%
%
\linethickness= 0.500pt
\setplotsymbol ({\thinlinefont .})
\plot 18.144 20.828 19.003 21.692 /
\linethickness= 0.500pt
\setplotsymbol ({\thinlinefont .})
%
%
\plot 16.669 19.325 	16.711 19.238
	16.751 19.165
	16.829 19.061
	16.913 19.004
	17.010 18.986
	17.131 19.016
	17.221 19.092
	17.290 19.201
	17.349 19.325
	17.408 19.450
	17.478 19.558
	17.568 19.635
	17.689 19.664
	17.810 19.635
	17.901 19.558
	17.970 19.450
	18.030 19.325
	18.089 19.201
	18.158 19.092
	18.249 19.016
	18.371 18.986
	18.492 19.016
	18.582 19.092
	18.651 19.201
	18.710 19.325
	18.769 19.450
	18.838 19.558
	18.929 19.635
	19.050 19.664
	19.171 19.635
	19.262 19.558
	19.331 19.450
	19.390 19.325
	19.449 19.201
	19.518 19.092
	19.608 19.016
	19.729 18.986
	19.851 19.016
	19.942 19.092
	20.011 19.201
	20.070 19.325
	20.130 19.450
	20.199 19.558
	20.290 19.635
	20.411 19.664
	20.532 19.635
	20.622 19.558
	20.692 19.450
	20.751 19.325
	20.810 19.201
	20.879 19.092
	20.969 19.016
	21.090 18.986
	21.206 19.032
	21.286 19.137
	21.354 19.251
	21.431 19.325
	21.431 19.325
	/
%
%
\linethickness= 0.500pt
\setplotsymbol ({\thinlinefont .})
\plot 19.022 21.702 20.235 20.491 /
%
%
\plot 20.011 20.626 20.235 20.491 20.100 20.716 /
%
%
%
\linethickness= 0.500pt
\setplotsymbol ({\thinlinefont .})
\plot 20.235 20.491 21.852 18.874 /
%
%
\plot 21.628 19.009 21.852 18.874 21.718 19.099 /
%
%
%
\linethickness= 0.500pt
\setplotsymbol ({\thinlinefont .})
\plot 21.852 18.874 22.593 18.133 /
%
%
\put{$\mu$} [lB] at 18.0 23.591
%
%
\put{$x_3$} [lB] at 19.033 24.591
%
%
\put{$x_2$} [lB] at 15.240 17.590
%
%
\put{$x_1$} [lB] at 22.670 17.590
%
%
\put{electron self-energy} [lB] at  3.4 16.669
%
%
\put{vertex correction} [lB] at 16.0 16.669
\linethickness=0pt
\putrectangle corners at  1.875 24.744 and 22.669 15.0
\endpicture}
\caption{One loop diagrams contributing to the
electron self-energy and the electron-electron-photon vertex
in QED \label{diagrams}}
\end{center}
\end{figure}
Let us first
calculate the electron self-energy, given by the first Feynman
graph in Fig~\ref{diagrams}. The bare expression is
\begin{equation}
  \Sigma(x) = e^2 \g_\alpha \prop(x) \dsl \prop(x) \g_\alpha ~,
\label{sigma}
\end{equation}
where $x=x_1-x_2$. Due to translation invariance,
$\Sigma$ only depends on $x$. Notice that
Eq.~(\ref{sigma}) involves no integration, in contradistinction
with the corresponding expression in momentum 
space\footnote{In
general, coordinate space calculations involve one integral
less than the momentum space ones. If one is interested in
quantities defined at  determined momenta
(like scattering amplitudes), the missing
integral must be done at the end as a Fourier transform
(without regulator!). One example when this Fourier transform is
not needed is the calculation of beta functions.}.  
After some straightforward (four-dimensional) diracology
and the use of Leibnitz rule to extract the derivative, one
obtains
\begin{eqnarray}
  \Sigma(x) & = & - e^2 \dsl (\prop(x))^2 \nonumber \\
            & = & - e^2 \dsl \B[1] ~. 
\end{eqnarray}
The renormalized value is, from Eq.~(\ref{renB}),
\begin{equation}
  \Sigma^R(x) = \frac{1}{64\pi^4} e^2 \dsl \Box
  \frac{\log x^2 M^2}{x^2} ~. 
\label{sigmar}
\end{equation}
Let us now deal with the vertex correction (see
Fig.~\ref{diagrams}). Reading
directly from the Feynman rules,
\begin{equation}
   V_\mu(x,y) = (-ie)^3 \g_\alpha \dsl^x \prop(x) \g_\mu
   (- \dsl^y) \prop(y)
   \g_\alpha \prop(x-y) ~,
\label{v1}
\end{equation}
with $x=x_1-x_3$ and $y=x_2-x_3$.
Simplifying the Dirac algebra and using systematically
the Leibnitz rule to rearrange derivatives,
$V_\mu(x,y)$ can be expressed in terms of triangular basic
functions:
\begin{eqnarray}
V_\mu(x,y) & = & ie^3 \{  -2 \g_b \g_\mu \g_a (
  \d_a^x \d_b^y \T[1] + \d_a^x \T[\d_b]
  - \d_b^y \T[\d_a])  \nonumber \\
  & & \mbox{} -2 \g_\mu \T[\Box] + 
  4 \g_a \T[\d_a\d_\mu] \}~.
\label{v2}
\end{eqnarray}
The renormalized expression is obtained directly from
Eqs. (\ref{renTbox}) and~(\ref{renTdd}):
\begin{eqnarray}
V_\mu^R(x,y) & = & ie^3 \{  -2 \g_b \g_\mu \g_a (
  \d_a^x \d_b^y \T[1] + \d_a^x \T[\d_b] \nonumber \\
  && \mbox{} - \d_b^y \T[\d_a])  
  + 4 \g_a \T[\d_a\d_\mu - 
  \frac{1}{4}\delta_{a\mu} \Box] \nonumber \\
  && \mbox{} - \frac{1}{4}
  \frac{1}{(4\pi^2)^2} \g_\mu \Box \frac{\log x^2 M^2}{x^2}
  \delta(x-y)
  - \frac{1}{8} \frac{1}{4 \pi^2} 
  \g_\mu \delta(x) \delta(y)\} ~.
\label{vr}
\end{eqnarray}
Once the graphs have been renormalized in coordinate
space, one can perform a Fourier transform (without any 
regulator)
to obtain the corresponding finite expressions in 
momentum space. 
We need the Fourier transforms of the basic
functions in Eqs. (\ref{sigmar}) and~(\ref{vr}).
The latter are more involved, so let us see in some
detail how to calculate them.
The Fourier transform of a distribution ``of
two variables'', $f(x,y)$, is
\begin{equation}
  \hat{f}(p,p') = \int {\mathrm d}^4 \! x \, e^{ix\cdot p}
  e^{iy \cdot p'} f(x,y)~.
\end{equation}
With the integration by parts prescription, total
derivatives in $f(x,y)$ yield
\begin{equation}
  \d_\mu^x \rightarrow -i p_\mu ~;~\d_\mu^y \rightarrow 
  -i p_\mu^\prime ~. 
\end{equation}
For the finite triangular functions we have:
\begin{eqnarray}
  \hat{\T}[\O] & = & \int {\mathrm d}^4 \! x {\mathrm d}^4 \! y \, 
  e^{ix\cdot p}
  e^{iy \cdot p'} \prop(x) \prop(y) \O^x \prop(x-y) \nonumber \\
  & = &  \int {\mathrm d}^4 \! x {\mathrm d}^4 \! y \, e^{ix\cdot p}
  e^{iy \cdot p'} \nonumber \\
  && \mbox{} \times \int \frac{{\mathrm d}^4 \! k_1}{(2\pi)^4} 
  \frac{{\mathrm d}^4 \! k_2}{(2\pi)^4} 
  \frac{{\mathrm d}^4 \! k}{(2\pi)^4} \,
  e^{-ix \cdot k_1} e^{-iy \cdot k_2} e^{-i(x-y) \cdot k}
  \frac{\hat{\O}(k)}{k_1^2 \, k_2^2 \, k^2} \nonumber \\
  & = & \int \frac{{\mathrm d}^4 \! k_1}{(2\pi)^4} 
  \frac{{\mathrm d}^4 \! k_2}{(2\pi)^4} 
  \frac{{\mathrm d}^4 \! k}{(2\pi)^4} \,
  \frac{\hat{\O}(k)}{k_1^2 \, k_2^2 \, k^2} \nonumber \\
  && \mbox{} \times (2\pi)^4 \delta(p-k_1-k) (2\pi)^4 \delta(p'-k_2+k) 
  \nonumber \\
  & = & \int \frac{{\mathrm d}^4 \! k}{(2\pi)^4} \,
  \frac{\hat{\O}(k)}{(p-k)^2 (p'+k)^2 k^2} ~,
\label{finiteFT}
\end{eqnarray}
where $\hat{\O}(k)$ is obtained from $\O^x$ by the
replacement $\d^x \rightarrow -ik$. 
The integrals in Eq.~(\ref{finiteFT}) appear 
(with a regularization which is not
present here) in standard one-loop calculations in momentum
space and can be evaluated
with standard techniques. We shall do that later on, in the limit
$p' \rightarrow 0$.
On the other hand, the Fourier transform of the renormalized basic 
function in Eq.~(\ref{vr}) reduces to 
Eq.~(\ref{DRFT}): 
\begin{eqnarray}
  \hat{\T}^R[\Box] & = & 
  \int {\mathrm d}^4 \!x {\mathrm d}^4 \!y
  \, e^{i x \cdot p} e^{i y \cdot p'}
  \frac{1}{64\pi^4} \Box \frac{\log x^2 M^2}{x^2} 
  \delta(x-y)  \nonumber \\
  &=& \int {\mathrm d}^4 \!x \, e^{i x \cdot (p+p')} 
  \frac{1}{64\pi^4} \Box \frac{\log x^2 M^2}{x^2} \nonumber \\
  & = & \frac{1}{16\pi^2} \log (\frac{(p+p')^2}{\bar{M}^2})~.
\end{eqnarray} 
With all these formulae, we get the renormalized 
vertex correction in momentum space: 
\begin{eqnarray}
  \hat{V_\mu}^R(p,p') & = & ie^3 \{-\g_b \g_\mu \g_a (
  -p_a p'_b \hat{\T}[1] - i p_a \hat{\T}[\d_b] + 
  i p_b' \hat{\T}[\d_a])  \nonumber \\
  && \mbox{}  + 4\g_a \hat{\T}[\d_a \d_\mu 
  - \frac{1}{4} \delta_{a\mu}\Box] \nonumber \\
  && \mbox{} - \frac{1}{16\pi^2} \g_\mu 
  ( \log(\frac{(p+p')^2}{\bar{M}^2}) + \frac{1}{2} ) \}~.
\label{momvr}
\end{eqnarray}
The Fourier transform (in one variable) of the self-energy
in Eq.~(\ref{sigmar}) is directly given by Eq.~(\ref{DRFT}):
\begin{equation}
  \hat{\Sigma}^R(p) = -i \frac{e^2}{16\pi^2} \not \! p 
  \log (\frac{p^2}{\bar{M}^2})~.
\label{momsigmar}
\end{equation}

\section{The vertex Ward identity}
Finally, let us verify the Ward identity between the renormalized 
vertex correction and electron self-energy. 
For 1PI Green functions it reads
\begin{equation}
(\d_\mu^x + \d_\mu^y) V_\mu^R(x,y) =
  i e \Sigma^R(x-y) (\delta(x) - \delta(y)) ~,
\label{WI}
\end{equation}
where $\d^{x_3} f(x_1-x_3,x_2-x_3) = - (\d^x + \d^y) f(x,y)$ 
has been used to express it
in the translated variables $x$ and $y$.
At points away from the origin this identity must
hold, since the bare Green functions
have not been modified there. A possible disagreement 
can only arise from the contact terms at $x=y=0$. In fact,
both sides of Eq.~(\ref{WI}) are distributions, and to compare
them one must either use formal properties of delta functions,
etc, or integrate with an arbitrary test function $\phi(x,y)$. 
In particular, one can  perform a Fourier transform
($\phi(x,y)=e^{ix\cdot p}e^{iy\cdot p'}$), which retains all the
information. In other words, we can check the Ward identity in
momentum space, 
\begin{equation}
  -i(p_\mu + p_\mu') \hat{V}_\mu^R(p,p') =
  ie [\hat{\Sigma}^R(-p') - \hat{\Sigma}^R(p)] ~,
\label{momWI}
\end{equation}
using the momentum space renormalized Green functions in 
Eqs. (\ref{momvr}) and~(\ref{momsigmar}).
For simplicity, we consider the
limit $p' \rightarrow 0$ ({\it i.e.\/}, the Fourier transform
in $y$ reduces to an integral without any weight). 
In this limit the relevant integrals
reduce to
\begin{eqnarray}
  \hat{\T}[\d_\alpha] & \stackrel{p'\rightarrow 0}{\rightarrow} &
  \frac{i}{16\pi^2} \frac{p_\alpha}{p^2} ~, \\
  \hat{\T}[\d_\alpha \d_\beta - \frac{1}{4} \delta_{\alpha\beta} \Box]
  & \stackrel{p'\rightarrow 0}{\rightarrow} &
  - \frac{1}{32\pi^2}
  \frac{p_\alpha p_\beta - \frac{1}{4} \delta_{\alpha\beta} p^2}{p^2}~,
\end{eqnarray}
whereas $\hat{\T}[1]$ is  logarithmically infrared divergent and
$p_\alpha' \hat{\T}[1] \stackrel{p'\rightarrow 0}{\rightarrow}0$. 
With these values, we obtain 
for both sides
of the Ward identity in Eq.~(\ref{momWI}), in the limit
$p' \rightarrow 0$, the same result:
\begin{displaymath}
    -\frac{e^3}{16\pi^2} \not \! p 
    \log \frac{p^2}{\bar{M}^2} ~.
\end{displaymath}
Since both sides are equal, 
the Ward identity is indeed satisfied: constrained DR
has preserved it automatically, {\it i.e.\/}, without
any {\it a posteriori\/} adjustment.
  
\section{Conclusions}
DR is a renormalization method which works in 
coordinate space and does not introduce any intermediate
regulator~\cite{FJL}. We have shown how it
can be easily applied to the calculation and
renormalization of one-loop Feynman diagrams. 
In many cases, worked out calculations at higher orders are 
also simpler in DR than in other methods~\cite{FJL,systematic}. 
Some nice features of the method are the following:
\begin{itemize}
\item It is minimal, in the sense that  
the Green functions are never modified except at the singular
points.
\item It does not change the space-time dimension, easing
the diracology and tensor manipulation, and the treatment
of chiral theories~\cite{chiral,WZ,g2}.
\item One integration less has to be performed, unless
one is interested in some quantity defined for fixed
external momenta. In such case, one has to Fourier transform the
renormalized expressions, but without any regularization.
\item Some overlapping divergencies disentangle in
coordinate space.
\item It is better suited for theories which are
naturally defined in coordinate space, like
theories with conformal invariance~\cite{nonabelian}, 
in curved space or at finite 
temperature~\cite{curved}.
\end{itemize}
In the constrained procedure of DR~\cite{CDR} 
all the local terms are fixed. This determines
a renormalization scheme which turns out to be symmetric
in all known examples~\cite{CDR,g2}. Here, we have used constrained DR 
to calculate two one-loop Green functions in QED and 
we have verified that the Ward identity relating 
them  is automatically fulfilled after renormalization. 
We have dealt with massless theories for simplicity. 
The treatment of massive theories in (constrained) DR is
worked out in Ref.~\cite{massiveDR} (Ref.~\cite{CDR}).
\section*{Acknowledgements}
It is a pleasure to thank A. Culatti and R. Mu\~noz Tapia for a 
fruitful collaboration when developing this subject.
This work has been supported by CICYT, contract number AEN96-1672,
and by Junta de Andaluc\'{\i}a, FQM101. MPV thanks MEC for financial
support.

\end{document}